# Late-Night Thoughts About the Significance of a Small Count of Nuclear or Particle Events

Ivan V. Aničin

*Faculty of Physics, University of Belgrade, Belgrade*
*Serbia and Montenegro*
e-mail: anicin@ff.bg.ac.yu

Reconciliation of frequentist and Bayesian approaches to elementary treatment of data in nuclear and particle physics is attempted. Unique procedure to express the significance of a small count in presence of background is henceforth proposed and discussed in some detail.

## 1. INTRODUCTION

It has been for more than fifty years now that nuclear and particle physicists worry about the unique way of expressing the significance of their conclusions which rely on small number of registered events of given signature, and especially so when some of these events are suspected to be of origin different than the one of current interest. This worry has in recent years eventually grown into a nightmare (hence the title of this paper), for new physics beyond Standard models is, if at all, necessarily represented by only a small signal immersed in the high background of the prevailing phenomena already embraced by the Standard models. From the early days of Regener (1951) [1], till PhyStat2003 [2], the physicists were exploring the rich heritage of the mathematicians in search of the final solution to the problem. Distressingly diversified body of literature on the subject has henceforth emerged, while great number of recent workshops and conferences devoted partly to this matter [3,4,5,6,7] speaks of the annoying situation. The results of all these efforts are accessible via the links found at the CDF Statistics Committee site [8]. Particle Data Group also recommends a number of possible approaches [9].

It is not our wish here to enter the endless dispute between the frequentists and the Bayesians about the fundamentals of probability theory. Our aim is a modest one; we shall at best try to give a somewhat different interpretation to common situations with counting statistics *only* (though Barlow emphasized that "everything is a counting experiment" [10]), including the important case when some phenomena other than the one under scrutiny are suspected to contribute to the overall number of indistinguishable counts. We shall reinterpret the key distribution function which occurs in both the frequentist and Bayesian approach to counting statistics and will hopefully be able to reconcile the two opposed views, at least when applied to elementary problems of counting statistics, and perhaps make everyday life easier for a practicing nuclear and particle physicist, as well as contribute to more homogeneous presentation of our results.

## 2. THE CASE OF A BACKGROUNDLESS COUNT – the one-parameter problem

Most of what follows is, of course, *deja vu*. We start with considering the inferences which we are allowed to make in the simplest of cases, when we have, in a given measurement time, counted a certain number of events of given signature, and when we either do not care about their origin or are 100% sure that they are all of the same origin, and when we believe that they satisfy the conditions required for their distribution to be Poissonian.

**2.a. Our knowledge of the average count after a single (small) count has been observed**

Consider first the case when we have counted nothing, what is, of course, how every experiment begins (and some even end). Let us recall that this eventless interval is exponentially distributed, and that on the average it lasts longer when the average counting rate is lower. If we assume that the process is Poissonian, such that the true, but prior to the experiment completely unknown average count in measurement time $\tau$ is $N$ (or the average counting rate $R$ is $N/\tau$), the (conditional) probability to obtain n counts in this measurement time is by assumption

$$P(n|N) = N^n e^{-N} / n!. \qquad (1)$$

According to the frequentist interpretation the probability of the zero count, if the average count was $N$, thus must have been $P(0|N) = e^{-N}$, and upon actually obtaining the zero count this distribution becomes all that we know about $N$ without further elaboration (we shall not consider here neither the alternative construction approach of Neyman, nor the so called unified approach, which are not easily compared with the Bayesian). We thus accept that, granting our initial assumptions were correct, this function represents the total information content of our result.

In the usual frequentist approach this function is obtained by thinking of the Poisson distribution (after the experiment has been performed, i.e. a single sample of it obtained) as being a function of $N$ (instead of $n$), and is called the likelihood function, the name being justified by the obvious conclusion that now we know that it is more likely for the average count to be close to zero than to have some higher value. The term "likelihood" is used instead of "probability" to reflect the fact that $N$ is not a stochastic variable, but is a parameter of the distribution which has a definite value, though at this instance not known to us exactly. This function is used for parameter estimation in the form of the so-called maximum likelihood method in the manner similar to the one we shall be using here. To the Bayesians this function is known as the posterior pdf, and is obtained by applying the Bayes' theorem to the Poissonian likelihood function (which by every criterion conforms to the frequentist definition of a pdf, and also note that the term "likelihood" now refers to a different function, what only adds to the confusion) which must be multiplied by an arbitrary constant, which is termed the uniform prior, and reflects our prior ignorance about the probability to obtain the zero count. Here then, this function is considered as a genuine probability distribution, minding that the Bayesian probability is, contrary to the frequentist probability which is meaningful for stochastic variables only, defined as a "degree of belief" that anything under consideration might have a certain value. The Bayesians thus find appropriate to integrate their posterior pdf (which is in this case formally identical to the frequentist likelihood) in order to obtain the probability (which is again a degree of belief) for the unknown parameter, the average count N, to be in a certain interval, what is the procedure to which the frequentists strongly object due to the non-stochastic character of *N*. The two parties therefore deal with formally the same function, but in two conceptually different ways. Formal identity of the two functions is of course well known. We just consider this to be more than a mere coincidence. If the two approaches did not have any crossing points we would be forced to conclude that one of them is altogether wrong, for two completely disparate views of the same thing cannot both be right (what is an occasionally overlooked truism).

Now that we have reviewed the basics of the two confronted views in this simplest of cases, we may try to bring them closer together by ascribing a slightly different meaning to the frequentist likelihood function, along the lines suggested in our introduction to the problem. We thus imply that it is our *knowledge* of the true average count that is specified by the above distribution. If we denote the degree of our knowledge of *N* in the light of the measured count n by *K(N|n)* then in the case of zero recorded counts $dK=K(N|0)dN=e^{-N}dN$. We thus imply that this function is meaningful to integrate, in order to **quantify our current knowledge about the average count being in a certain interval** (as it is in the Bayesian tradition, where it is interpreted as a pdf, but in the Bayesian sense of probability, the "degree of belief" then corresponding to our "knowledge"). For these purposes the function even need not be normalized. Our knowledge about the average count is in this situation thus concentrated around zero; we are 90% convinced that the average count is smaller than 2.3:

$$\int_0^{2.3} K(N|0)dN = 0.9 \; ; \qquad (2)$$

we are only 10% convinced that it is greater than 2.3, etc., and, quite uselessly, we *know for sure* (what corresponds to the norm of the function) that the average count has a certain value between zero and infinity. Since the two parties interpret (and name) this function differently, and since we suggest that in both approaches it actually represents our current knowledge of the true average count in the light of the available data, to stress that in our view this function is neither the probability distribution of average counts *N*, nor the likelihood function in the frequentist sense ("likelihood", according to the consulted dictionaries (Webster, Oxford, etc.) is anyway only an alternative word for probability, or probability in disguise) from now on we call it the *"**knowledge density function**",* or "kdf" for short. The interval within which given percentage of our conviction about the value of the average count resides, we shall, in good frequentist tradition, still call the confidence interval and the confidence level (CL) respectively. The degree of arbitrariness as to how to position the confidence interval, at a given confidence level, remains. This is still a matter of convention. It is perhaps only plausible to always choose the narrowest one at a given CL, which necessarily contains the maximum of the kdf.

(A comment is perhaps appropriate here. If we were to prolong the measurement, and if we were still left with the null result, the kdf of the average count will remain the same, but our knowledge of the average counting *rate* (which is always our final objective) will improve. For a non-zero result repetition of identical measurements is good for control purposes, but is in every other respect identical to a single measurement performed during the same overall measurement time.)

For an arbitrary recorded count *n*, the kdf is

$$K(N|n) = N^n e^{-N} / n!. \qquad (3)$$

This is the Gamma distribution, which, together with our knowledge of *N*, peaks at *n*, and has both the dispersion and the mean (which is of no consequence here) equal to *n*+1 (loosely, this Poissonian kdf we also call the inverse Poissonian). For any non-zero *n* it correctly vanishes at zero, since once any non-zero count is obtained we know for sure that the average count cannot be zero. Also, what is usually disregarded, this kdf is at the same time the probability density function of time intervals (since $N=\tau R$) between (*n*+1) scaled counts. This is good, for it allows the treatment of the results of preset time and preset count measurements on equal footing.

There are now two possibilities to analyze this kdf in order to express the degree of our current knowledge of the average count. We may either integrate it in order to quantify our knowledge about the average count being in a given interval, or alternatively, we can make use of the approach adopted by the maximum likelihood method, which analyzes the logarithm of a kdf rather than the kdf itself. Integrating the kdf is straightforward, while the analysis of ln(kdf) calls for some comments. Our knowledge of the previously unknown parameter is concentrated around the maximum of both the kdf and *ln*(kdf), and sharpness of both functions determines the size of the confidence intervals for the parameter. The choice of the logarithm of the kdf, instead of the kdf itself, is motivated mostly by the ease with which the limits of confidence intervals are derived in this case. As it turns out (and is seen clearly when the kdf is normal, or is approximated by the second term of Taylor expansion around the maximum) the limits of confidence intervals at confidence levels which correspond to those of *s* standard deviations for the normal distribution, are determined by the intersections of the lines of equal *lnK*-values which are obtained when $s^2/2$ is subtracted from the value of the *ln*(kdf) in its maximum, $lnK_{max}$. Thus, the 1σ interval, or the confidence interval at the 68% confidence level, is the one which is enclosed between the intersection points of the ln(kdf) curve with the iso-*lnK* line which is obtained when 0.5 is subtracted from $lnK_{max}$, the 2σ, or the 95.4% confidence level interval, is the one within the intersections of the iso-*lnK* line which is by 2 lower than the maximum, for the 3σ level it is 4.5, we are 90% convinced that the true average count is enclosed within the intersections with the iso-*lnK* line which is 1.35 lower than the maximum *lnK* value (*s*=1.64), etc.

To illustrate how the two possibilities operate let us have a closer look at the *n*=3 case. In Fig.1a the kdf *K*(*N*,*n*=3) as given by Eq.(3) is presented, while Fig.1b presents the ln*K*(*N*,*n*=3) function. In Fig.1b the ln$K_{max}$−$s^2/2$ lines together with their intersections with the ln*K* function are marked, and in Fig.1a the corresponding points bear the same letters. The integrals between these limits are found to be only insignificantly smaller than the confidence levels of 68% (1 sigma), 95.4% (2 sigma), and 99.73% (3 sigma), what justifies the use of both methods, even for small counts. It is interesting to note that the confidence interval at the 68% confidence level, which is equal to 3.41, though asymmetric around the measured *n*=3 count, is only insignificantly different from the 2√3=3.46 (symmetric) common wisdom interval. This difference gets bigger at higher confidence levels. It is seen the at higher confidence levels the confidence intervals soon become uselessly large and that their lower limits asymptotically approach zero.

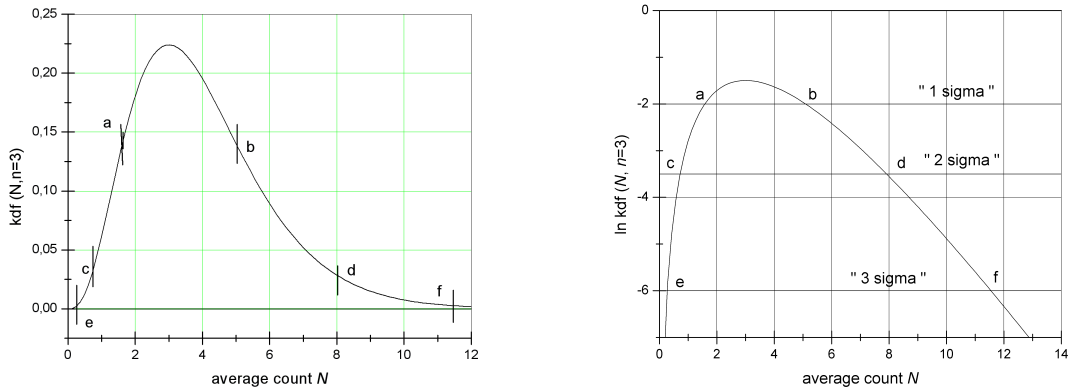

**Fig. 1**. a) The kdf K(N,n=3), as given by Eq.(3), and b) the lnK(N,n=3) function. In both figures the limits of 1σ, 2σ and 3σ confidence intervals are marked.

## 2.b. At high count inverse Gaussian is a good approximation to Poissonian kdf

Another important property of the kdf is that, as the Gauss distribution of counts *n* for a given average count *N* becomes better approximation to the Poisson as the average count increases, so does the inverse Gauss distribution, or the Gauss kdf for the average count *N*, as given by our Eq.(4):

$$K(N|n) = \frac{1}{\sqrt{2\pi N}} \exp\left[-\frac{(N-n)^2}{2N}\right] \qquad (4)$$

become better approximation of the inverse Poisson, or the Gamma function as given by our Eq.(3), as the recorded count *n* increases. To illustrate this point, in Fig.2 we present the two distributions for the observed count *n*=3 and *n*=20. In a way, this makes the approach we advocate here nicely consistent, and speaks in its favor. Fig.2b also reveals that at high counts both kdfs at low confidence levels become symmetric and normal-like.

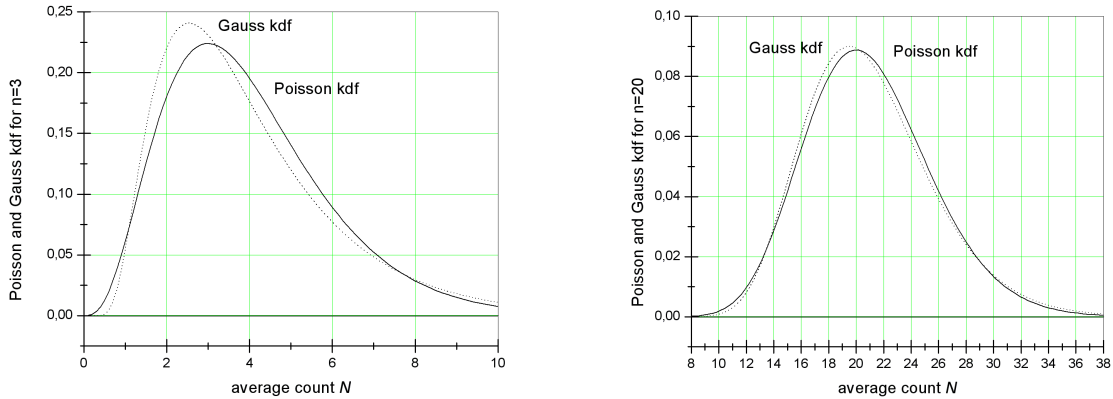

**Fig. 2**. Comparison of the Poisson kdf [Eq(3)], and Gauss kdf [Eq.(4)], for the observed count a) *n*=3, and b) *n*=20. The bigger the count, the difference between the two gets smaller, like the difference between the Poisson and Gauss distribution of counts for a given average count gets smaller with the increasing average count.

**2.c. Predicting the next count**

The knowledge of the average count as represented by the kdf, which is the final result which we seek, may be used for another seemingly important purpose, which would further justify its meaning and name which we have suggested here. In accord with the main goal of all of science, which is to predict the future on the basis of our current knowledge of the past, we might be interested in predicting the next count $k$, after we have previously observed a single count $n$. Only if we knew the average count $N$ exactly (which is the knowledge achieved only after an infinite number of samplings) would our prediction (or expectation) of the next count $k$ be distributed by the Poisson distribution $P(k|N)$. However, after a single observed count $n$ our knowledge of $N$ is not sharp but is distributed as $K(N|n)$ and a contribution of any possible $N$ to the probability of the occurrence of the next count $k$ is $P(k|N) \times K(N|n)$. The overall probability to obtain a given count $k$ after count $n$ has been observed is thus equal to:

$$p(k|n) = \int_0^\infty P(k|N) \times K(N|n)\, dN. \qquad (5)$$

This function, which we shall name the "prediction distribution function" (another pdf!) can be found explicitly:

$$p(k|n) = \int_0^\infty \frac{N^k e^{-N}}{k!} \frac{N^n e^{-N}}{n!} dN = \frac{(k+n)!}{k!n!} 2^{-(k+n+1)} \qquad (6)$$

This pdf is properly normalized, its mean is $n+1$, and it favors higher values significantly more than the Poisson distribution. The difference between the Poisson distribution, which would be a valid prediction function if we knew for sure that the average count is $N=5.0$, and this pdf for the observed count $n=5$, is illustrated in Figure 3. Though this function is not of great practical use, and we shall here make no further mention of it, it is conceptually interesting as it demonstrates the full meaning of statistical determinism in counting experiments.

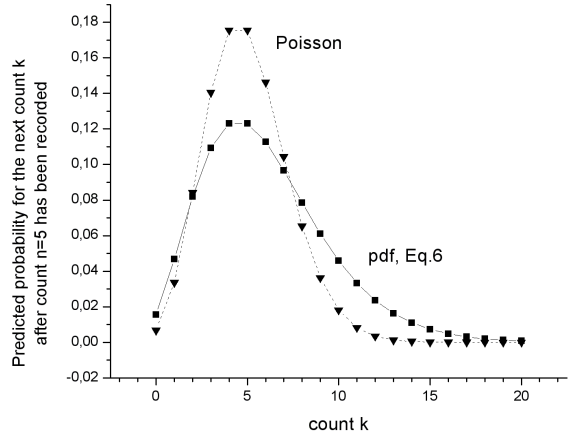

**Fig.4**. Comparison of the prediction distribution function for the next count k when the count *n*=5 has been previously observed, as defined by our Eq. (6), and the Poisson distribution, which is the prediction function for the count k when it is exactly known that the average count is *N*=5.0.

We believe that we have thus arrived at a heuristic proof that the two approaches are, at least in the case of elementary counting statistics, equivalent, assuming our interpretation of the frequentist likelihood function is adopted by the frequentists, and provided the Bayesians adopt the uniform prior as the only correct one, for any other prior will simply yield the wrong kdf.

### 3. THE CASE OF A COUNT WITH BACKGROUND – the two-parameter problem

Now that we have, under the conditions stated above, hopefully agreed upon the equivalence of frequentist and Bayesian views in the simplest one-parameter case, we may in the same spirit proceed with the analysis of realistic cases of small number of counts, in the presence of background, what is a typical two-parameter problem most frequently met with in both the nuclear and particle physics.

**3.a. The algorithm**

To this end we shall deal in some detail with the procedure which is usually pursued by the Bayesians for the analysis of weak spectral lines in gamma-ray spectra situated on the continuous constant background, which may, under our terms, be adopted by the frequentists as well. The procedure is nicely presented by Sivia [11], and was recently used by Klapdor et. al. [12] to support their first ever hint of a positive result for the neutrinoless double beta decay. As we shall see it will turn out to be quite useful generally, for the definition of significance of arbitrary count in the presence of background which is known with arbitrary precision.

We assume that the spectral data consist of a certain number of channels $x_i$ ($i=1,2,...,m$) with counts $n_i$ which are distributed as

$$P(n_i|N_i) = (N_i)^{n_i} e^{-N_i} / n_i!, \qquad (7)$$

with average values $N_i$ which are in turn supposed to satisfy the equation

$$N_i(A,B) = C\{A\ exp[-(x_i-x_0)^2/2w^2] + B\}. \qquad (8)$$

The position $x_0$ and the width $w$ of the Gaussian spectral line are supposed to be known in advance with negligible uncertainty, while the height of the line $A$ (or its intensity) is the parameter of interest, and the height of the constant spectral background B is a nuisance parameter which must be determined only in order to find the value of $A$ (the value of $C$ is of no consequence here). This situation is common in high resolution gamma-ray spectroscopy, and has been as such in practically the same way analyzed in real life by Klapdor, ref.12. Assuming that the counts in the channels are independent, the whole set of data $\{n_i\}$ now has the probability to appear, or (Bayesian) likelihood, equal to the product of probabilities for individual channel counts:

$$P[\{n_i\}|N_i(A,B)]=\prod_i P[n_i|N_i(A,B)]. \qquad (9)$$

Now, considered as a function of unknown parameters $A$ and $B$ this is at the same time equal to what we have called a knowledge density function, or both to the frequentist likelihood function and the Bayesian posterior pdf obtained with a constant prior:

$$K(A,B|\{n_i\})=P[\{n_i\}|N_i(A,B)]. \qquad (10)$$

According to the interpretation of both approaches, as agreed above, this function determines completely our knowledge about the parameters $A$ and $B$ in the light of the measured spectrum $\{n_i\}$.

Following the tradition of both the frequentist maximum likelihood approach to parameter estimation, and the Bayesian analysis of the posterior pdf, we shall also analyze not the kdf itself, but its logarithm, which is in this case equal to:

$$ln[K(A,B|\{n_i\})] = ln\ P[\{n_i\}|N_i(A,B)] = const + \sum_i \{n_i\ ln[N_i(A,B)] - N_i(A,B)\}. \quad (11)$$

As in the one-parameter case, our knowledge of the parameters is concentrated around the maximum of both the kdf and $ln$(kdf), and sharpness of both functions determines the sizes of the confidence intervals for the parameters. One important difference from the one-parameter case is now that, at a certain confidence level, possible values of the parameter necessarily become negative, or unphysical. The question is then only at what confidence level, $CL_{max}$, this takes place. We are then at maximum $CL_{max}$ % convinced that our signal is non-zero. Thus, if the $1\sigma$ interval, or the confidence interval at the 68% confidence level, is the last one that does not include zero, we are only 68% convinced that the result is non-zero, and so on. The $5\sigma$ level, which is near our complete conviction that it contains the true value of the average count, and which is nowadays recommended for the results of extreme importance (like the Higgs), requires that the large confidence interval which is bound within the iso-$lnK$ line $s^2/2=12.5$ below the maximum, does not comprise zero. Some subtleties we shall discuss later on, in the examples.

The two-parameter problem is a convenient one for it can be solved graphically. Minding the accuracy of our statistical inferences at low statistics (or statistical errors of our statistical errors, which hardly justify more than a single significant figure in the quantification of our knowledge), the accuracy of the graphical method is quite appropriate. Insisting on higher accuracy would anyway be in a somewhat bad taste. The easiest way is to find numerically, and then inspect graphically, the values of $lnK(A,B)$ for a sufficiently fine grid of $A$ and $B$ values around the maximum, so as to cover the wanted confidence interval. This is best illustrated by an example.

**3.b. Weak Gaussian spectral line on a low constant background**

We first analyze in some detail a typical germanium gamma-ray spectrum in the close surrounding of an expected spectral line. In the first numerical example the counts in the consecutive 11 channels are: 1,2,0,1,4,3,2,1,0,2,0. The Gaussian line of unit width, $w=1$, is expected in channel $x_0=6$. The spectral background is low and the line is, if there is any, weak. Upon inspecting the numbers we do not expect the background to be higher than, say, 1, and the line to be higher than 3, what would be the anticipated coordinates ($A_{max}, B_{max}$) of the maximum of $lnK$. Since the line is weak we are most probably interested in the maximum confidence level at which it may be considered existent, and this is why we would like to know the values of $lnK$ for $A=0$, so we include zero in the range of values for $A$. On the other hand, we might want to know the confidence interval for the intensity at a certain confidence level, which is why we have to extend the range of values for $A$ well past the maximum of $lnK$, perhaps up to 6 in this case.

Though we are not interested in the value of background *B*, we still have to determine the interval for its values, where we shall investigate the values of *lnK*. This is somewhat tricky because the two parameters, *A* and *B*, are in principle correlated, most probably in such a way that to the positive deviation of one parameter from its average there corresponds a negative deviation of the second one from its average (what is called anticorrelation). Correlation between the parameters is judged by the tilt of symmetry axes of the iso-*lnK* contours projected onto the *A*,*B* plane. It may thus happen, as we shall see, that the confidence limits at a given confidence level for *A* correspond to the values of *B* well away from its value at $lnK_{max}$. This is difficult to predict, and if this graphical method is adopted, one often has to work in a number of steps, until all wanted information is obtained. We shall take this interval somewhat wider that it might appear necessary, say from zero to 2. The steps of 0.1 in both parameters are sufficiently fine, and a small program will easily find the values of [*A*, *B*, *lnK*(*A*,*B*)] in such a grid.

If the resulting three columns are imported into ORIGIN®, and converted to matrix form, one among the Plot3D routines will draw a contour plot of wanted iso-*lnK* lines projected onto the (*A*,*B*) plane. The result of this procedure for our example is presented in Fig.4.

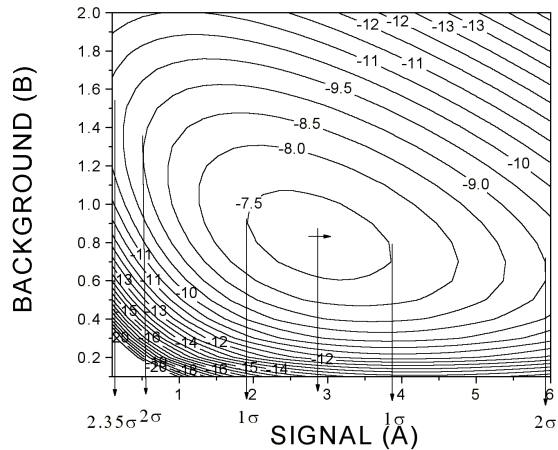

**Fig.4.** Analysis of the kdf for the spectrum 1,2,0,1,4,3,2,1,0,2,0, assuming a Gaussian line one channel wide in channel 6, and flat background elsewhere.

The iso-*lnK* lines are drawn at 0.5 intervals, starting from $lnK_{max}$. The coordinates of the maximum of the kdf are: $A_{max} \approx 2.8$ and $B_{max} \approx 0.8$. The limits of the 1σ and 2σ intervals are denoted in the figure, and it may be seen that they correspond to positions of vertical tangents (for background these are the horizontal tangents) to the $lnK_{max}-0.5$ and $lnK_{max}-2$ iso-*lnK* lines. The aforementioned effects of (anti)correlation between *A* and *B* may also now be inspected. We may thus at the 68%CL quote the result as $A \approx 2.8^{+1.0}_{-1.0}$, or at the 95.4%CL as $A \approx 2.8^{+3.0}_{-2.4}$, etc.

Alternatively, we may say, for instance, that we are 95.4% convinced that the height of the supposed line is somewhere in the interval from 0.5 to 6. Another possibility is to express the confidence level at which we are still convinced that there is a line at all. To find this confidence level we have to find the last iso-$lnK$ level which lies completely in the region of positive $A$ values, which is the one that touches the vertical axis. If we have determined the confidence level which corresponds to this iso-$lnK$ line to be $(CL)_{max}$, then the confidence level (expressed in %) at which we are still convinced that there is a line at all is $(CL)_{max}+[100-(CL)_{max}]/2$ (what is equivalent to integrating the kdf). In our case the iso-$lnK$ line which touches the vertical axis is the one which is $2.35\sigma$ below $lnK_{max}$, so that $(CL)_{max}$ is 98%. We are thus 99% convinced that there is a line at all. Finally, at the $3\sigma$ level we may not claim the line any more, since the $lnK_{max}-4.5$ level has the left vertical tangent well into the region of negative $A$ values.

We see that every positive result for the signal can always be interpreted in two basically different ways. First, it can be expressed as a two-sided confidence interval at a given confidence level (or, what is equivalent, as a definite value with definite errors, so as to encompass the same interval). Secondly, it can be expressed, always at the higher confidence level than in the first case, as a maximum level at which we are convinced that the result may be considered non-zero (what is sometimes called the one-sided interval). Which of the possibilities we shall use depends primarily on the value of the confidence level, $(CL)_{max}$, at which the two ways of expressing the result start to differ. If the highest confidence level at which the two-sided confidence interval can be still stated is too low, smaller than one sigma for instance, then we shall rather state the maximum level at which the signal can still be considered non-zero.

When the maximum of the kdf occurs for the negative value of the signal, what we shall illustrate in the examples which follow, we can only state the maximum confidence level (which is always smaller than 50%) at which the signal can still be considered positive.

**3.c. "Single-point Gaussian line" on a low constant background**

We now examine the transitional case between the spectral analysis, which we performed above, and the simple event counting case, which is of more interest in particle physics. We consider the same spectral situation as before, but with the width of the Gaussian line degenerated to a single channel (what is the infamous case of a "single-point Gaussian", which occurs at low dispersions). The width of such a "line" may be taken anything smaller than or equal to, say, $w=0.1$. If other channels contain only background, the problem is then equivalent to testing the hypothesis that this single count belongs to the background or not, what amounts to determination of the confidence level at which it may be considered a fluctuation of the background, or that at which it may not. Consider the following sequence of counts: 0,1,3,0,2,7,1,2,0,2,2, and suppose that there is a "single-point Gaussian" in channel 6, where the count is 7. Same procedure as above now yields the result which is presented in Fig.5.

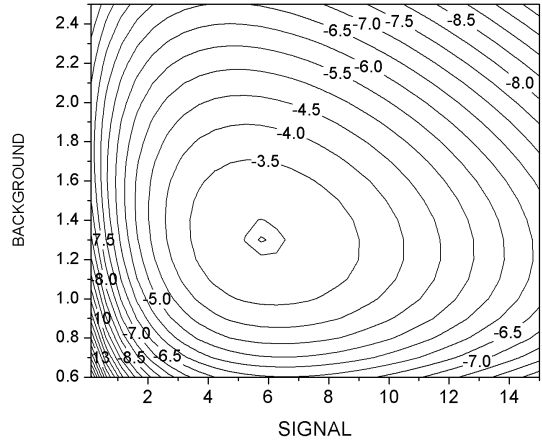

**Fig.5**. Analysis of the kdf for the spectrum 0,1,3,0,2,7,1,2,0,2,2 assuming a "single-point Gaussian" line 0.1 channel wide in channel 6, and flat background elsewhere.

We see that our knowledge of average background now peaks around 1.3 and the height of the "line" therefore peaks around the remaining 5.7, and that this can be considered a "line" even at the 3σ confidence level, but not higher.

**3.d. Small signal+background count on a small background count**

Next example will take us to the case of a simple counting problem. If we now let not only the line, but also the background to degenerate to a single channel, then we are left with two counts only (one of which is supposed to be background and the other signal+background), and the problem of finding out the significance of the difference between them. Our algorithm will, without any changes, provide us with the answer. Let the two counts be 3 and 6. Let us first suppose that background is 3 and background plus signal is 6. We have to remember that all the counts are Poissonian, so that we expect the kdf for background to be inverse Poisson, while that for the signal will be something different, but with the dispersion which is the sum of dispersions of both counts. The result of the procedure which is performed under these assumptions is presented in Fig.6.

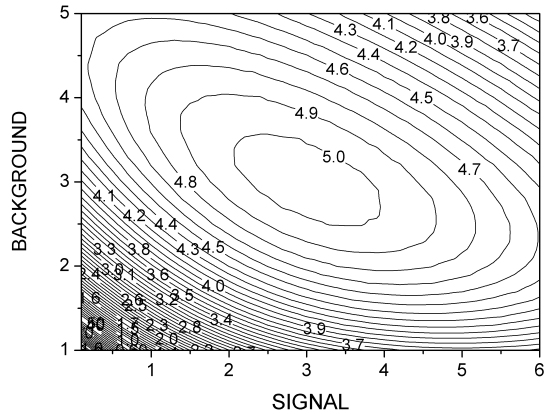

**Fig.6.** Analysis of the kdf for the two counts only, 3 and 6, first of which is supposed to be Poissonian background, and second signal+background. The iso-*lnK* lines now differ by 0.1.

It is seen that the kdf behaves as expected. The signal may be considered significant at the level of 1σ only, what amounts to the statement that we are at maximum 68% convinced that count 6 belongs to a different population than count 3.

Next we invert the situation. We keep the same counts as in the previous example, but exchange their roles; we now suppose that predicted background is Poissonian 6 and that signal+background turned out in the experiment to be 3. Result of the analysis is presented in Fig.7.

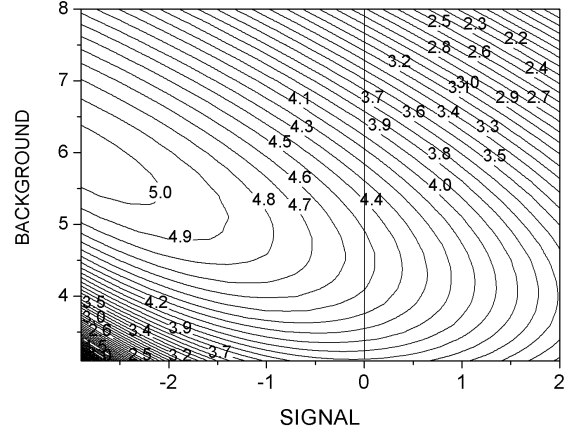

**Fig.7.** Analysis of the kdf for counts 3 and 6, as above, but with the roles of the counts interchanged.

We see that the signal becomes positive at the confidence level around 70%, so that we may be about 30% convinced that the signal is still non-zero.

### 3.e. Small signal+background count on a small "non-Poissonian" background count

Our algorithm is easily adapted to deal with the cases when background count in a given measurement time is in advance predicted with any given precision (is "non-Poissonian"), either from separate measurement of the duration different than that of the actual experiment which measures signal+background, or from an appropriate Monte Carlo simulation. Since in our algorithm only the average value of counts in the channels which are declared as background determines its properties, irrespective of the actual distribution of counts in these channels, we are entitled to structure the background in such a way so as to satisfy our needs. Refering to our example of subsection 3.c. this can be done in the following way.

If in the experiment which is supposed to measure signal+background the predicted background at the 68%CL is $n_B \pm \delta_B$, and if $\delta_B = f \sqrt{(n_B)}$ (with $f \leq 1$), then we construct the background so as to be distributed in $\eta = 1/f^2 = \text{CINT}(n_B/\delta_B^2)$ channels, which contain integer counts which all add up to $\text{CINT}(\eta n_B)$, where CINT denotes operation of taking the closest integer of the expression in the parentheses (we have accepted a certain inconsistency here - a symmetric confidence interval at a given confidence level for the background estimate, instead of an asymmetric one - but the difference is quite small at low CL, and absolute rigor is out of place). The way in which we shall distribute these counts is arbitrary. To illustrate how this works let us work out an example.

Suppose the predicted background at the 68%CL in a certain experiment is 3.2±0.5. If it was a Poissonian prediction its dispersion would have been 3.2, and half of the 1σ confidence interval would be √3.2=1.79, instead of 0.5 in our case. The confidence interval is thus by the factor $1/f$=1.79/0.5=3.6 narrower than it would have been if the background were determined on the basis of the measurement of the same duration as that of the actual experiment. That means that the duration of the background measurement is $\eta=1/f^2$ = CINT($n_B/\delta_B^2$)=13 times longer than that of the signal+background measurement, and that during that time CINT($\eta n_B$)=41 background counts must have been observed (or simulated). To construct the background with such properties for the purposes of our analysis we thus have to distribute 41 counts in 13 channels, in an arbitrary way. Let us take this to be 12×3+5, or: 3,3,3,3,3,3,3,3,3,3,3,3,5. If in the actual experiment 6 counts have been observed, which we potentially attribute to signal+background, we then have to analyze the following "spectrum": 3,3,3,3,3,3,3,3,3,3,3,3,5,6, assuming the existence of a "single-point line" in channel 14, and considering all the rest as background. When we do this we obtain the result presented in Fig.8.

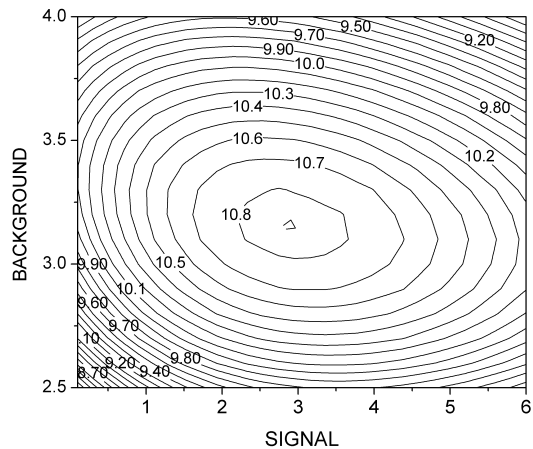

**Fig.8.** Analysis of the kdf for the counting experiment in which the expected background at 68%CL is 3.2(5), which is represented by the counts 3,3,3,3,3,3,3,3,3,3,3,3,5 in background channels, and the signal+background measured count is 6, which is represented by the count in a separate channel, where a single-point Gaussian line 0.1 channels wide is supposed to exist.

We see that the characteristics of the background are as required, and that increase of its accuracy over the Poissonian accuracy resulted in the increased confidence level at which the signal may be considered non-zero, as compared to our example in Fig. 6 (though perhaps not as great an increase as might have been expected).

**3.f. High signal+background and background counts**

Finally, when the counts in the channels get bigger than twenty, or maybe even ten, both the count pdf and its inverse, the kdf, get close to normal, and it becomes irrelevant which algorithm one adopts. The results of our analysis then become equal to those of the standard least squares analysis, which in principle cannot be applied when the counts are Poissonian (though, as evidenced by the discussion of our Fig.2b, at low confidence levels the LSF method, even at low counts, produces virtually the same results, at higher confidence levels, where the asymmetry of the distributions is more apparent, they start to differ significantly from those of the method applied here). To demonstrate this we apply our algorithm to the counting experiment where the expected background is Poissonian 10 and the signal+background count is 30. We expect the confidence interval for the signal at the 68%CL to be $2\sqrt{(30+10)}$ wide, and in Fig.9, where this situation is analyzed assuming the existence of the single-point line in the channel which contains signal+background, we see that it is indeed so.

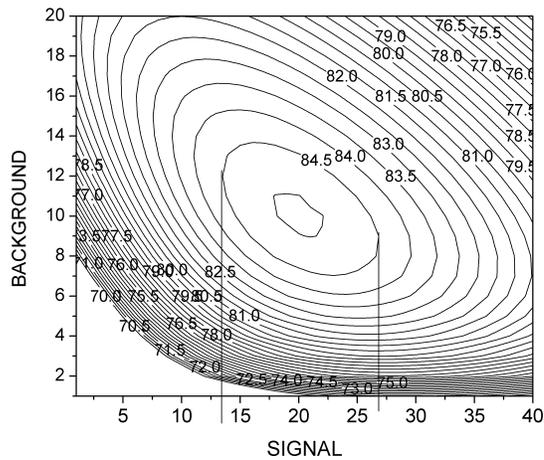

**Fig.9.** Analysis of the kdf for the two counts, 10 and 30, first of which is supposed to be Poissonian background, and second signal+background. The iso-*lnK* lines differ by 0.5.

To conclude, we believe that the examples, which were chosen to be representative of elementary counting experiments in nuclear and particle physics, demonstrate that the significance of results from these experiments may be meaningfully assessed by always using one and the same algorithm, which we have here elaborated and, hopefully, justified.

This work is partly supported by the Serbian Ministry of Science and Environment, under Projects No.1451 and No.1859.

**References:**

1. V.H.Regener: "Statistical Significance of Small Samples of Cosmic-Ray Counts", Phys. Rev. **84**(1951)161
2. Louis Lyons: "PHYSTAT2003", Nucl. Instr. and Meth. in Phys. Res. **A534**(2004)15
3. F.James, L.Lyons, Y.Perrin, Eds.: "Workshop on Confidence Limits", CERN 17-18 January 2000, CERN Yellow Report 2000-005, Geneva 2000, available on-line at: http://ph-dep.web.cern.ch/ph-dep/Events/CLW/papers.html
4. "Workshop on Confidence Limits", Fermilab, 27-28 March 2000 http://conferences.fnal.gov/cl2k/fredjames_lectures.pdf
5. "Advanced Statistical Techniques in Particle Physics" Conference, Durham 2002, http://www.ippp.dur.ac.uk/Workshops/02/statistics/proceedings.shtml
6. "Conference on Statistical Problems in Particle Physics, Astrophysics and Cosmology", PhyStat2003, Stanford 2003 http://www.slac.stanford.edu/econf/C030908/proceedings.html
7. PHYSTAT "Workshop on Statistical Software for Physics and Astronomy", March 1-2, 2004 Michigan State University http://user.pa.msu.edu/linnemann/public/workshop/program.htm
8. http://www-cdf.fnal.gov/physics/statistics/statistics_home.html and http://www.nevis.columbia.edu/~evans/references/statistics.html
9. Particle Data Group, Phys. Lett. **B592**, 1(2004), and http://pdg.lbl.gov
10. R. Barlow, Introduction to Statistical Issues in Particle Physics, Presented at PHYSTAT 2003, ref.5
11. D.S.Sivia: "Data Analysis – A Bayesian Tutorial", Clarendon, Oxford 1996
12. H.V. Klapdor-Kleingrothaus, A. Dietz, H.V. Harney and I.V.Krivosheina, "Evidence for Neutrinoless Double Beta Decay" Modern Physics Letters **A 16**, No. 37 (2001) 2409-2420